\documentclass[]{spie}  %>>> use for US letter paper
\usepackage{etoolbox}
\csundef{ver@cite.sty}
\usepackage{amsmath,amsfonts,amssymb}
\usepackage{graphicx}
\usepackage[colorlinks=true, allcolors=black]{hyperref}
\usepackage{gensymb}
\usepackage{wrapfig}
\usepackage{orcidlink}
\usepackage{subcaption}
\usepackage{nopageno}
\usepackage[backend=bibtex,style=numeric,sorting=none,natbib=false]{biblatex}
\title{Across the soft gamma-ray regime: utilizing simultaneous detections in the Compton Spectrometer and Imager (COSI) and the Background and Transient Observer (BTO) to understand astrophysical transients}

\author[a,b, *, \orcidlink{0000-0002-0699-2544}]{Hannah C. Gulick}
\author[c, d]{Eliza Neights}
\author[b]{Samer Al Nussirat}
\author[b]{Claire Tianyi Chen}
\author[b]{Kaylie Ching}
\author[b]{Cassandra Dove}
\author[b]{Alyson Joens}
\author[d]{Carolyn Kierans}
\author[b]{Hubert Liu}
\author[d, e]{Israel Martinez}
\author[b]{Tomas Mician}
\author[b]{Shunsaku Nagasawa}
\author[b]{Shreya Nandyala}
\author[b]{Isabel Schmidtke}
\author[b]{Derek Shah}
\author[b]{Andreas Zoglauer}
\author[f]{Kazuhiro Nakasawa}
\author[g, h]{Tadayuki Takahashi}
\author[b]{Juan-Carlos Martinez Oliveros}
\author[b, \orcidlink{0000-0001-5506-9855}]{John A. Tomsick}

\affil[a]{University of California, Berkeley, Department of Astronomy, Berkeley, CA, United States, 94720, USA}
\affil[b]{Space Sciences Laboratory, University of California, Berkeley, CA 94720, USA}
\affil[c]{George Washington University, USA}
\affil[d]{NASA Goddard Space Flight Center, 8800 Greenbelt Road, Greenbelt, MD 20771, USA}
\affil[e]{University of Maryland, USA}
\affil[f]{KMI, Nagoya University, Nagoya, Aichi, Japan}
\affil[g]{Kavli Institute for the Physics and Mathematics of the Universe (WPI), University of Tokyo, Kashiwa, Chiba 277-8583, Japan}
\affil[h]{Department of Physics, University of Tokyo, Bunkyo, Tokyo 113-0033, Japan}

\authorinfo{Send correspondence to H.C.G or J.T.: \\H.C.G. E-mail: hannah\_gulick@berkley.edu \\ J.T. E-mail: jtomsick@berkeley.edu}

\begin{document} 
\maketitle

\begin{abstract}
The Compton Spectrometer and Imager (COSI) is a NASA funded Small Explorer (SMEX) mission slated to launch in 2027. COSI will house a wide-field gamma-ray telescope designed to survey the entire sky in the 0.2-5 MeV range. Using germanium detectors, the instrument will provide imaging, spectroscopy, and polarimetry of astrophysical sources with excellent energy resolution and degree-scale localization capabilities. In addition to the main instrument, COSI will fly with a student collaboration project known as the Background and Transient Observer (BTO). BTO will extend the COSI bandpass to energies lower than 200 keV, thus enabling spectral analysis across the shared band of 30 keV–2 MeV range. The BTO instrument will consist of two NaI scintillators and student-designed readout electronics. Using spectral information from both the COSI and BTO instruments, physics such as the energy peak turnover in gamma-ray bursts, the characteristics of magnetar flares, and the event frequency of a range of transient phenomena will be constrained. In this paper, we present the expected science returnables from BTO and comment on the shared returnables from the COSI and BTO missions. We include simulations of gamma-ray bursts, magnetar giant flares, and terrestrial gamma-ray flashes using BTO's spectral response. Additionally, we estimate BTO's gamma-ray burst detection rate and find that BTO will detect $\sim$100--150 gamma-ray bursts per year, with at least 10\% of the events being sGRB.
\end{abstract}

\keywords{Gamma-ray Survey, Gamma-ray Transients, COSI, Time Domain, Multimessenger}

\section{INTRODUCTION}
\label{sec:intro}  % \label{} allows reference to this section

%The transient gamma-ray sky holds insights into the extreme energetics which power the highest energy events in the Universe as well as on Earth.  

The transient gamma-ray sky holds insights into the extreme energetics which power and map events such as black hole accretion \cite{Abdo2010, Ajello2020}, electron-positron annihilation \cite{Bouchet2010, Knodlseder2005}, compact object formation and evolution \cite{Abbott2017a, Campana2006, Kumar2015, MacFadyen1999}, and the progenitors of multimessenger events \cite{Abbott2017a, Abbott2017b, Eichler1989}. Transient gamma-ray events are important for understanding broader topics such as galaxy formation and evolution, the history of star formation in the Milky Way, and dark matter. With peak energies ranging from $\sim$1--100 keV for gamma-ray burst (GRB) and magnetar flares \cite{Peer2006, Kaspi2017}, to 100s of MeV for blazars \cite{Ajello2020, DeAngelis2021}, the gamma-ray bandpass is extensive. As a consequence, a single instrument is not capable of observing across the entire gamma-ray regime.
%Instead, gamma-ray telescopes must have specialized instrumentation enabling observations in a specific energy range.
%and even GeV for GRB associated with neutron star--neutron star mergers \cite{Adriani2018}, 

Spectral signatures from gamma-ray transients are variable across the gamma-ray bandpass, requiring multiple instruments to observe and constrain event characteristics.
%Signatures from gamma-ray transients are not constant across the gamma-ray bandpass. 
Events such as magnetar flares or Galactic electron-positron annihilation occupy a finite fraction of the gamma-ray bandpass, peaking at energies of 60 keV \cite{Kaspi2017} and 511 keV \cite{Knodlseder2005, Bouchet2010}, respectively. Other phenomenon such as the energy peak turnover in GRBs exhibit energy dependent structure \cite{Peer2006, Band1993}, and thus require observations across a range of energies to observe different underlying emission mechanisms. Therefore, to properly constrain the physics powering the most energetic phenomenon in the Universe, simultaneous observations are needed from complimentary instruments specialized to different energy ranges. 

Historically, gamma-ray observatories have included multiple instruments on a single spacecraft platform to enable observations across a wide range of energies. For example, the INTErnational Gamma-Ray Astrophysics Laboratory (INTEGRAL) which launched in 2002 included both a gamma-ray imager \cite[IBIS]{Lebrun2003} and spectrometer \cite[SPI]{Vedrenne2003} covering energies from $\sim$20 keV to $\sim$10 MeV. These instruments were complimented by the Joint European X-ray Monitors \cite[JEM-X 1 \& 2]{Lund2003} which provided images in the 3--35 keV bandpass. The Astro-rivelatore Gamma a Immagini Leggero \cite[AGILE]{Marisaldi2010, Marisaldi2015} included two imagers---the Gamma Ray Imaging Detector \cite[GRID]{Chen2013} sensitive in the 30 MeV--50 GeV energy ranges as well as SuperAgile \cite[SA]{Feroci2007} sensitive in the 18--60 keV energy range. Additionally, the \textit{Fermi} satellite includes both the Large Area Telescope \cite[LAT]{Atwood2009} providing images in the 20 MeV to 300 GeV range as well as the Gamma-ray Burst Monitor \cite[GBM]{Meegan2009} providing spectra in the 8 keV to 40 MeV.

The future Compton Spectrometer and Imager \cite[COSI]{tomsick2023} mission slated to launch in 2027, will also include multiple instruments covering a wide energy bandpass. These instruments include the main COSI gamma-ray telescope and a second soft gamma-ray spectrometer known as the Background and Transient Observer (BTO). While COSI will provide high resolution images, spectroscopy, and polarization measurements in the 200 keV to 5 MeV range, BTO will extend the COSI bandpass to lower energies by providing spectral information in the 30 keV to 2 MeV range. Due to their overlapping bandpasses as well as shard FOVs, COSI and BTO will enable spectroscopic analysis of gamma-ray transients over a large gamma-ray bandpass. Future observations made with COSI and BTO will help to isolate GRB photospheric and non-thermal emission components \cite{Ryde2010, Guiriec2011, Peer2006} and constrain the energy peak turnover which typically occurs below 200 keV \cite{Band1993}. Additionally, BTO's lower bandpass and wider FOV is better suited to detect magnetar flares \cite{Kaspi2017} and will thus improve the mission's ability to probe the characteristic timescales, temperatures, and brightness profiles associated with these events. Finally, the combined COSI and BTO systems will yield observations of multimessenger events which will be used to localize the event progenitor and to measure the delay time between gravitational waves and their corresponding electromagnetic counterparts \cite{Abbott2017a, Abbott2017b}.

In this paper, we outline the full mission overview in Section \ref{sec:mission_overview} and include instrument specifics for COSI and BTO in Sections \ref{subsec:COSI} and Section \ref{subsec:BTO}, respectively. Section \ref{sec:simulations} reports on transient event simulations as well as the expected GRB event rates in BTO. The scientific advantages achieved through launching the complimentary COSI and BTO systems are also highlighted through simulated spectra of transient events in \ref{subsec:GRB}. Conclusions are made in Section \ref{sec:conclusions}.

\section{Mission Overview}
\label{sec:mission_overview}

The Compton Spectrometer and Imager (COSI) is a space-based gamma-ray survey mission that will provide imaging, spectroscopy, and polarimetry of astrophysical sources associated with the birth and death of stars, the formation of chemical elements in the Milky Way, and the progenitors of gravitational waves. Selected in 2021 as a NASA Small Explorer (SMEX) mission, COSI is slated to launch to a low-Earth-orbit (LEO) in 2027. As of April 2024, COSI passed its Key Decision Point review and was confirmed to proceed to Phase C. The COSI instrument is discussed further in Section \ref{subsec:COSI} and a full instrument outline is included in \cite{tomsick2023}.

In addition to the main instrument, the COSI SMEX proposal included a secondary proposal for a student collaboration project which was successfully funded. The student collaboration project aims to develop an additional gamma-ray detector system to fly on COSI---known as the Background and Transient Observer (BTO) (see Section \ref{subsec:BTO}). BTO will fly onboard the COSI spacecraft as a do-no-harm payload to provide observations at lower energies (30 keV to 2 MeV) and across a larger field-of-view (FoV) ($>$ 60\% of the sky) than the main COSI instrument. BTO's main science goals will be to detect gamma-ray transients and to constrain the soft gamma-ray background rates at different locations in the COSI orbit. Therefore, BTO will be highly complimentary to the COSI mission as a whole by extending the mission's bandpass, increasing the on-sky coverage for transient monitoring, and improving data analysis through improved background rate measurements.

%In accordance with the BTO mission goals, the project is fully student led and staffed---from payload manager to software, mechanical, and systems engineers. The BTO team would like to recognize the amazing mentorship and contributions from scientists, engineers, and advisors at the Space Sciences Laboratory, Kavli Institute for the Physics and Mathematics of the Universe, University of Tokyo, and Nagoya University. With this mentorship, the project has passed the COSI Preliminary Design and Key Decision Point reviews.

Along with COSI, BTO has passed the Preliminary Design and Key Decision Point reviews. BTO has a functioning engineering model currently.

\subsection{The Compton Spectrometer and Imager}\label{subsec:COSI}

\begin{figure} [ht]
   \begin{center}
   \begin{tabular}{c} %% tabular useful for creating an array of images 
   \includegraphics[height=8cm]{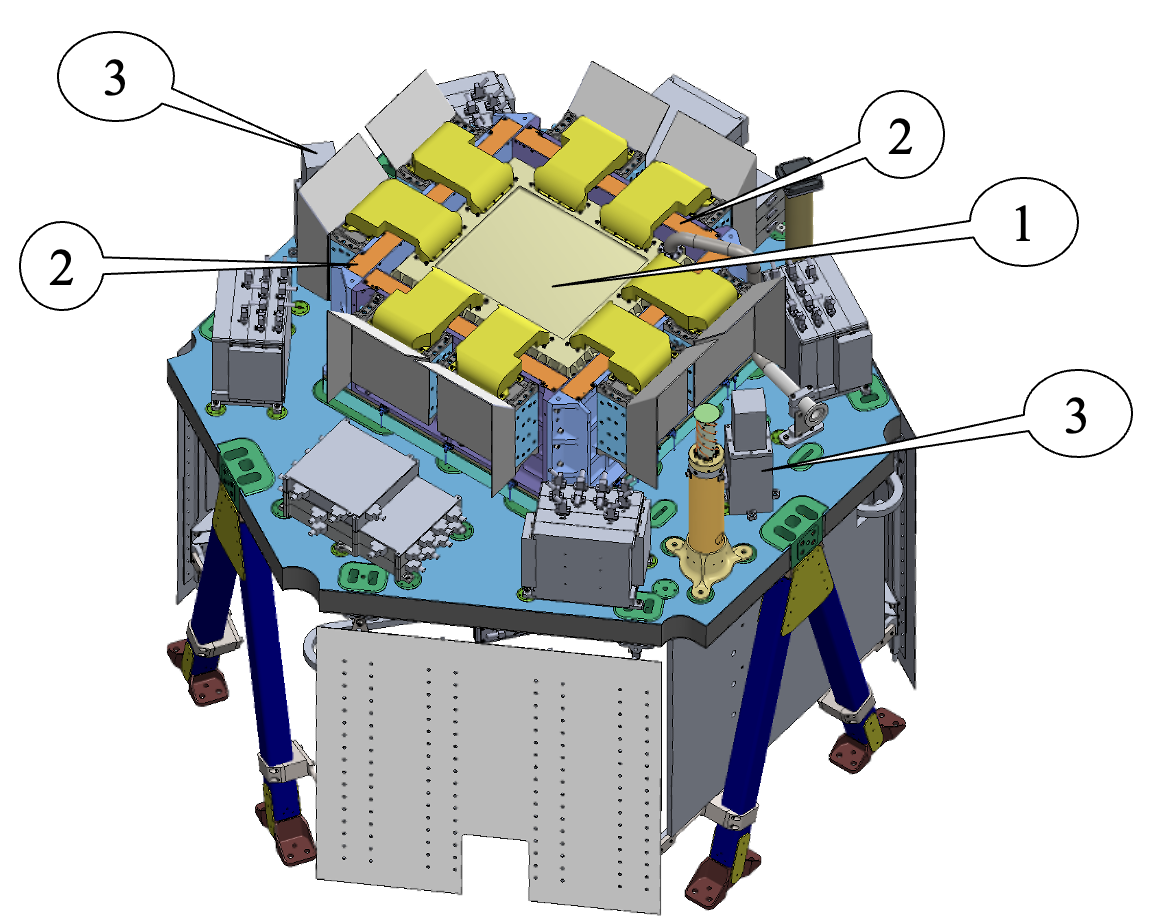}
   \end{tabular}
   \end{center}
   \caption 
   { \label{fig:PIP} 
A schematic of the COSI spacecraft. An in-depth description of the COSI instrument design is included in \cite{tomsick2023}. For the purposes of this paper, the following components are labeled: 1. the COSI germanium detectors contained in a vacuum cryostat, 2. the active BGO shields, and 3. the two BTO detectors.}
\end{figure}

COSI is a Compton telescope capable of observing from 200 keV to 5 MeV with an instantaneous FOV covering 25\% of the sky. Figure \ref{fig:PIP} shows the COSI spacecraft with the photon-sensitive subsystems labeled. The main telescope (labeled 1) is composed of cryogenically-cooled germanium detectors which use a double-sided strip structure to enable measurements of an interacting photon's x and y location as well as its depth (z). The instrument includes four stacks of four detectors (16 detectors total), each with 64 strips per side and a dimension of 8 cm $\times$ 8 cm $\times$ 1.5 cm. The detectors are surrounded by a guard ring electrode which performs signal vetoing. An Application Specific Integrated Circuit (ASIC) is used to read out the COSI detectors and guard ring. With the cryogenic cooler, the detectors are cooled to 80--90 K, allowing the instrument to achieve an excellent energy resolution of of 6 keV at 0.511 MeV and 9 keV at 1.157 MeV \cite{tomsick2023}. These resolutions enable unprecedented access to the low-energy part of the gamma-ray regime---otherwise known as the 'MeV Gap' due to the observational challenges associated with this bandpass (i.e. high background levels) and thus the lack of existing observations.

The germanium detectors are surrounded on four sides and underneath by active bismuth germanium oxide (BGO) scintillators. The BGO shields, shown in orange in Figure \ref{fig:PIP}, are readout by silicon photomultipliers (SiPMs) and act to provide passive background attenuation as well as anti-coincidence information on background radiation and gamma-rays that escape from the germanium detectors. Light curve information from the BGO shields will be downlinked from the spacecraft and used to improve background reduction in the COSI data as well as identify transient events.

Due to COSI's Compton telescope design, both the total energy and incident angle of the incoming photon can be identified and used to localize and reconstruct the source's image on the sky. COSI will be capable of imaging events with an angular resolution of 4.1$\degree$ and 2.1$\degree$ at 0.511 MeV and 1.809 MeV, respectively. As part of the mission requirements, COSI will create full Galaxy images at 0.511 MeV as well as four nuclear line energies. Additional information on COSI line sensitivities is included in Table 1 of \cite{tomsick2023}. Additionally, since the azimuthal component of the Compton scattering angle measured in the germanium detectors depends on the polarization of the incoming gamma-ray, COSI will also provide polarization measurements. The COSI detectors will be sensitive to polarization down to a flux limit of 1.4 $\times 10^{-10}$ erg cm$^{-2}$ s$^{-1}$ in the 0.2-0.5 MeV range.

With the use of these three observing capabilities, COSI will address four main science goals: 1. uncovering the origin of Galactic positrons through studies of the 0.511 MeV emission from antimatter annihilation in the Galaxy \cite{Bouchet2010, Knodlseder2005}, 2. revealing Galactic element formation by mapping radioactive elements from nucleosynthesis \cite{Eichler1989}, 3. gain insight into extreme environments through polarization measurements of accreting black holes \cite{Ajello2020}, and 4. probing the physics of multimessenger events through the observation of short gamma-ray bursts (sGRB) \cite{Abbott2017a, Abbott2017b}.

\subsection{The Background and Transient Observer}\label{subsec:BTO}

\begin{figure} [ht]
   \begin{center}
   \begin{tabular}{c} %% tabular useful for creating an array of images 
   \includegraphics[height=8cm]{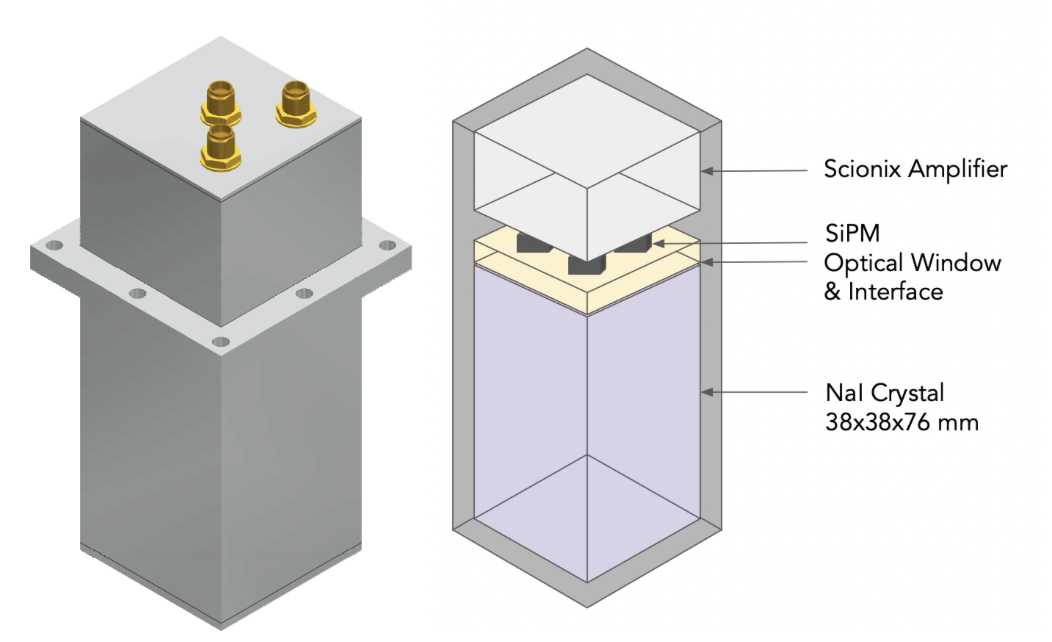}
   \end{tabular}
   \end{center}
   \caption 
   { \label{fig:BTO_det} 
A schematic of the Scionix NaI detector package customized for the BTO mission. }
\end{figure}

The BTO instrument will measure gamma-rays from transient events and monitor the soft gamma-ray background rates in the 30 keV to 2 MeV regime. BTO will utilize two detector modules---each composed of a rectangular NaI detector (see Figure \ref{fig:BTO_det}), an analog board with analog-to-digital converter (ADC), and a digital board. The BTO detectors are shown in Figure \ref{fig:PIP} on the outer edges of the spacecraft instrument deck---otherwise known as the Payload Interface Plate (PIP). The BTO electronics are housed in a single, stacked electronics box which is located on the underside of the PIP and therefore not visible in Figure \ref{fig:PIP}.

A trade study was conducted to optimize the scintillator material, size, and shape, including an analysis of afterglow signatures in CsI versus NaI scintillators Gulick et al. submitted. The custom BTO detectors are supplied by Scionix and utilize components with space heritage. The detectors contain a 3.8 cm $\times$ 3.8 cm $\times$ 7.6 cm NaI crystal hermetically sealed in a 4.5 cm $\times$ 4.5 cm $\times$ 12 cm aluminum housing. The detector energy resolution is 15--20\% at 662 keV. Figure \ref{fig:BTO_det} shows a diagram of the detector. Each detector includes four ArrayJ-60035-4P-PCB SiPMs for detector readout and a proprietary built-in amplifier. The flight model detectors include a single SMA input for $\pm$ 12 V power and two SMA outputs---one for the 'raw' SiPM signal and one for the amplified detector signal.

The detectors are located on the outer edges of the COSI PIP (see Figure \ref{fig:PIP}) to optimize the BTO FOV to $>$60\% of the sky. With only two detectors, BTO will not be capable of providing precise source localization. Therefore, the mechanical housing which affixes the detector to the PIP was designed with a 'pedestal' to raise the crystal by 11.5 cm. The pedestal---along with the detector placement---ensures the BTO detectors are not obscured by other components on the COSI PIP and results in the full COSI FOV being overlapped by part of each BTO detector's FOV. Thus any BTO detected event which occurs in the shared FOV will be localized to $\sim$4.1$\degree$ by the COSI germanium detectors. Furthermore, the placement of the BTO detectors on opposite sides of the COSI PIP allows shadowing from the germanium detectors, BGO shields, and other components on the COSI PIP to be used to roughly constrain the location of events outside of the COSI FOV to be of the Sun, Earth, or astrophysical origins.

Each detector will have its own set of readout electronics, consisting of an analog, ADC, and digital board. The analog board will process the amplified signal from the NaI detector and convert it to a digital signal with an analog-to-digital converter (ADC). Following the ADC, the digital board will be equipped with two modes of operation. Due to data downlinking limits, BTO will nominally operate in a 'binned mode' during which all counts over a 1 ms timescale will be saved to a binned histogram and downlinked. To preserve timing information for true gamma-ray events, however, the digital board will switch to an 'event-by-event mode' once triggered. In event-by-event mode, the timing and energy information for every event will be individually saved and downlinked. The event-by-event mode is especially important for the analysis of short gamma-ray events such as terrestrial gamma-ray flashes (TGFs).

Following the readout and processing system will be a power board and interface board. The power board will be responsible for converting and filtering the spacecraft's 28 V input to $\pm$12 V for the detector and analog/ADC boards and to $\pm$5 V for the digital boards. The interface board will be responsible for relaying commands and data to and from the BTO system to the spacecraft control panel. As a do-no-harm payload, the power and interface boards will act as the only electrical points of contact with the COSI spacecraft.

By observing at lower energies than COSI, BTO will probe interesting science associated with gamma-ray bursts (GRB) and magnetar flares. The BTO bandpass is ideal for constraining the energy peak turnover in lGRB and sGRB \cite{Band1993} and for isolating photospheric components in GRB spectra \cite{Peer2006, Ryde2010, Guiriec2011}. BTO will also be capable of measuring the time delay between gravitational wave events and their sGRB electromagnetic counterparts \cite{Abbott2017a, Abbott2017b}. Additionally, good timing resolution ($<$0.5 ms) is necessary to detect TGF events from lightening on Earth with timescales $<$ 1 ms) \cite{Smith2005, Grefenstette2009}. BTO's event-by-event mode achieves this resolution. Furthermore, due to the anti-coincidence requirement between the Compton telescope and active BGO shields on COSI, BTO will be the only instrument on the spacecraft which saves and downlinks TGF event information.

\section{Observing Astrophysical Transients with BTO \& COSI}\label{sec:simulations}

In this section we provide an in-depth discussion of the estimated science returnables from the BTO instrument for several transient phenomena. We use the Medium Energy Gamma-ray Library \cite[\textbf{MEGAlib}]{Zoglauer2006} to simulate a BTO detector response curve and corresponding event rates and spectra for GRBs, magnetar GFs, and TGFs. 

We also comment on the scientific advantages enabled through the compatibility of the COSI and BTO instruments. Due to their overlapping bandpasses and FOVs, the combined COSI and BTO instruments will provide spectral information from 30 keV to 2 MeV for gamma-ray transients which occur in their shared FOVs. The COSI response values and simulated light curves utilized in the sections below were created as part of a larger project which will be published in an upcoming manuscript by Neights et al. in prep.

\subsection{Gamma-ray Bursts}\label{subsec:GRB}

\subsubsection{GRB Detection Rates in BTO}

In this section, we estimate the number of GRBs detectable by BTO. For lGRB detection estimates, we use the \textbf{MEGAlib} package \textit{GRBSampler} which is well equipped to handle the larger lGRB population. This package uses a catalog of real GRB observed by the Burst and Transient Source Experiment (BATSE) \cite{Goldstein2013} over 3323 days (effective exposure time of 2390 days). The catalog contains 2,145 GRB, with 22\% of the events being sGRB. %1529 lGRB and 442 sGRB. 
To create a population of simulated GRB, the \textit{GRBSampler} package randomly samples probability weighted histograms of the following parameters: GRB duration (s), GRB average flux (photons/cm$^2$/s), the Band function lower photon index, the Band function upper photon index, the Band function break energy, and the GRB's polar ($\theta$) and azimuthal ($\phi$) angle on the sky. Note here that the average flux value given by MEGAlib is sufficient to estimate the detection significance for lGRB events, however, the peak flux is needed for sGRB events due to their peaked nature. Therefore, we include a different detection estimate method for sGRB below.

We simulated 5000 GRB to create a robust sample. To account for the Earth occultating the BTO FOV, only GRB with a $\theta$ $>$112.5$^\circ$ were included in the study. For each remaining GRB, we calculate the signal-to-noise ratio (SNR) via:

\begin{equation*}
    SNR = \frac{f_{GRB} \sqrt{t}}{\sqrt{f_{GRB}+f_{b}}}
\end{equation*}

\noindent
where $f_{GRB}$ is the GRB's flux (photon/s), $f_b$ is the background flux (photon/s), and $t$ is the exposure time (s). The background flux is calculated for the BTO geometry using \textbf{MEGAlib} as outlined in Gulick et al. submitted. The average expected background rate for a single BTO detector is 57 counts/s for an orbit with an 550 km altitude and 0$\degree$ inclination.

\begin{figure}
\begin{center}
\begin{tabular}{c}
\includegraphics[height=9cm]{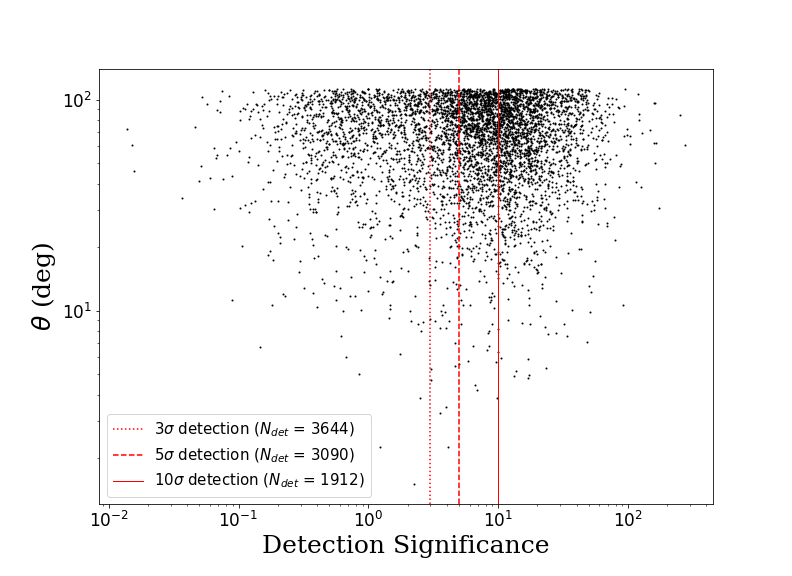}
\end{tabular}
\end{center}
\caption 
{ \label{fig:GRB}
The detection significance versus polar angle on the sky for 5000 simulated GRB events. In these simulations, a single BTO NaI detector geometry was used to calculate the expected flux from each GRB. The SNR is then calculated for each GRB as described in Section \ref{subsec:GRB}. The red dotted, dashed, and solid lines mark the 3, 5, and 10 $\sigma$ confidence levels. The variable $N_{det}$ gives the number of GRB which are detected above each significance level.} 
\end{figure}

Figure \ref{fig:GRB} shows the detection significance versus polar angle for the 5000 GRB simulated with the NaI detector. The 3$\sigma$, 5$\sigma$, and 10$\sigma$ levels are denoted by red dotted, dashed, and solid lines respectively. In the legend, $N_{det}$ refers to the number of GRB detected above each SNR interval.

To determine the number of GRB visible in BTO over the nominal mission duration of 2 years, we scale the simulated population with the BATSE detection rate. The BATSE catalog effective exposure time is 2390 days with 2,145 total detections. Extrapolating outwards to determine the effective exposure time required to detect the simulated data population, we expect BATSE to make 5000 detections in $\sim$4600 days or $\sim$12.6 years. Therefore, we scale the total simulated/observable population by 0.16 to estimate the number of detections over two years. Furthermore, the simulations were performed for a single BTO detector with an all-sky FOV so the data are further scaled by 0.6 to represent the BTO detector's FOV. 

This corresponds to $\sim$150 lGRB detections with both BTO detectors per year (295 GRB over the full 2-year mission) or a detection rate of $\sim$0.4 GRB per day. Comparatively, the Gamma-ray Burst Monitor on the \textit{Fermi} Observatory triggers on $\sim$200 lGRB per year \cite{Meegan2009}. GBM, while using the same NaI scintillation material, has a larger effective area ($\sim$120 cm$^2$ for a single detector \cite{Tsvetkova2022, Meegan2009}) than BTO (peak $\sim$50 cm$^2$ for both detectors) and observes from $\sim$8 keV to $\sim$40 MeV. Therefore, from the ratio of effective areas, it is expected that GBM has $\sim$2.4$\times$ the detection rate of BTO. Therefore, the current GRB rate estimates for BTO might be slightly overestimated (i.e. closer to 100 GRB per year). This value will be refined in future work as the BTO detector response matrix is completed across the full sky.

\begin{figure}
\begin{center}
\begin{tabular}{c}
\includegraphics[height=7cm]{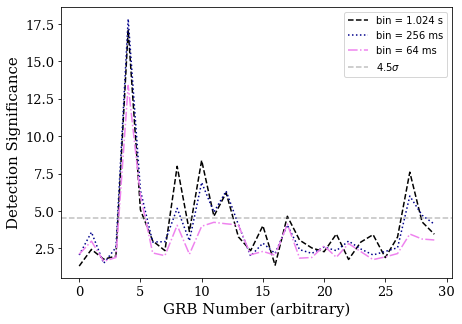}
\end{tabular}
\end{center}
\caption 
{ \label{fig:sGRB_detections}
The detection significance in a single BTO detector for all 30 sGRB detected by GBM in 2016. The sGRB lightcurves are binned into 64 ms (pink, dot-dashed line), 256 ms (dark blue, dotted line), and 1.024 s (black, dashed line) time bins. The 4.5$\sigma$ detection threshold is indicated by the grey dashed line.} 
\end{figure}

To estimate the number of sGRB detections, we use real lightcurves from GBM to calculate the detection significance at the GRBs peak count rate (i.e. instead of average flux). To do this, we download lightcurve data from the GBM burst catalog \cite{vonKienlin2020, Gruber2014, vonKienlin2014, Bhat2016}. We then bin these lightcurves into 64 ms, 256 ms, and 1.024 s time bins and calculate the detection significance in the peak time bin for each binning scheme. We choose to analyze the sGRB (t$_{90}$ $\leq$ 2 s) data from 01-01-2016 to 12-31-2016 to provide an estimate for the number of sGRB detected per year by GBM. This results in a population of 30 sGRB detections. We scale each GRB lightcurve by the effective area for a single GBM NaI and single BTO NaI detector to produce an lightcurve with a count rate reflective of the BTO effective area. 

Figure \ref{fig:sGRB_detections} shows the detection significance in BTO for all 30 sGRB events. For time bin sizes of 64 ms, 256 ms, and 1.024 s, a total of 2, 8, and 8 sGRB are detectable above 4.5$\sigma$ per year, respectively. This result agrees with the expected value derived by scaling the BTO and GBM effective areas for a single NaI detector ($\sim$10 sGRB/year). Furthermore, this improves the BTO sGRB to lGRB detection ratio making it closer to 8\%--10\%.

\begin{figure}
\begin{center}
\begin{tabular}{c}
\includegraphics[height=9cm]{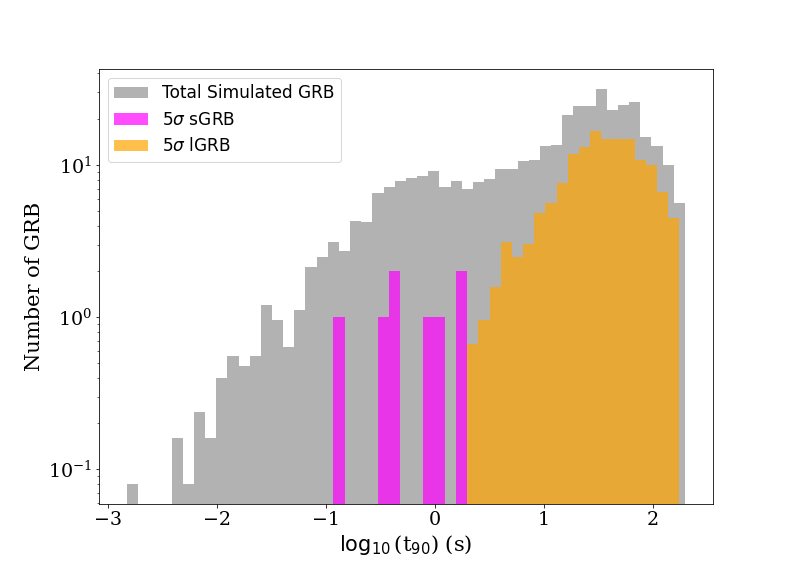}
\end{tabular}
\end{center}
\caption 
{ \label{fig:GRB_distrib}
%The $t_{90}$ distribution for GRB events simulated with \textbf{MEGAlib}. The histogram has been scaled from the initial simulated population containing 5000 events to reflect a 2-year observing period and an FOV covering $\sim$60\% of the sky (see Section \ref{subsec:GRB}). The $t_{90}$ distribution for all simulated events occurring over the 2-year mission window is shown in grey. The sGRB and lGRB events that are observable by BTO with a $>$5$\sigma$ detection level are shown in pink and orange, respectively. The distribution is fit with a bimodal function.
The $t_{90}$ distribution for GRB events simulated with \textbf{MEGAlib} (grey and orange) and downloaded from the GBM burst catalog (magenta). The MEGAlib histograms have been scaled from the initial simulated population containing 5000 events to reflect a 1-year observing period and an FOV covering $\sim$60\% of the sky (see Section \ref{subsec:GRB}). The GBM population was calculated over 1 year, and therefore not scaled. The $t_{90}$ distribution for all simulated events occurring over the 2-year mission window is shown in grey. The sGRB and lGRB events that are observable by BTO with a $>$5$\sigma$ detection level are shown in pink and orange, respectively. The discontinuous binning in the sGRB distribution is due to the both low number statistics and the face that it is real data reflective of the GBM timing resolution. } 
\end{figure}

Figure \ref{fig:GRB_distrib} shows the distribution of GRB event durations for all simulated GRBs in a single year in grey. The lGRB with a $>$5$\sigma$ detection level are shown in orange and sum to approximately 150 detections per year. At the 3$\sigma$ confidence level, this value increases to $\sim$170 lGRB per year. The sGRB detected with $>$5$\sigma$ confidence are shown in magenta and sum to 8 events per year. Typically, $\sim$20\% of GRB detections are sGRB, as shown by Konus-Wind and GBM \cite{Tsvetkova2022}. The Swift-Burst Alert Telescope (Swift-BAT) had a slightly lower rate with $\sim$9\% of detections contributed by sGRB. However, BTO's sGRB detection ratio of 8\%--10\% is slightly low. This may be due to the possible overestimate of lGRB detections as highlighted in the anaysis

%Similar to the results found in \cite{Lien2016} for the Swift-BAT telescope, we find that shorter gamma-ray events are difficult for the BTO geometry to detect if they are 1. highly off-axis or 2. both low flux and short duration. As is expected, the BTO sensitivity increases with exposure time thus BTO is capable of detecting high-flux sGRB, or low-flux sGRB with longer durations. Therefore, the BTO detectors will be biased toward the longest or brightest sGRB events. %To satisfy BTO science requirements, 7 sGRB and 100 lGRB must be detected during BTO's mission lifetime. At the 3$\sigma$ level, both of these requirements are met. Above 3$\sigma$ the detection rates for sGRB are too low to satisfy BTO's science.

%we expect that the lower and harder total count. This is because sGRB have characteristically harder energy profiles and the tall, thin BTO detector geometry is less suited to these higher energies. Therefore, only sGRB which occur along the longest axis of the BTO detector are confidently detected. 

\subsubsection{Simulated COSI \& BTO GRB Spectra}

\begin{figure}
\centering
\begin{subfigure}[b]{0.85\textwidth}
r   \includegraphics[width=1\linewidth]{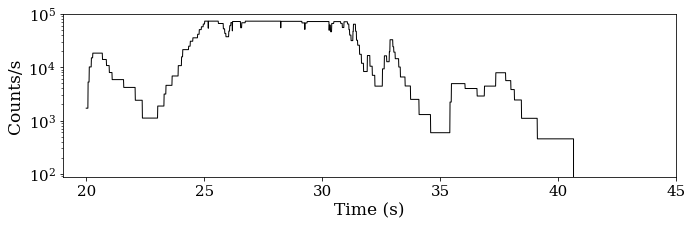} 
\end{subfigure}

\begin{subfigure}[b]{0.85\textwidth}
   \includegraphics[width=1\linewidth]{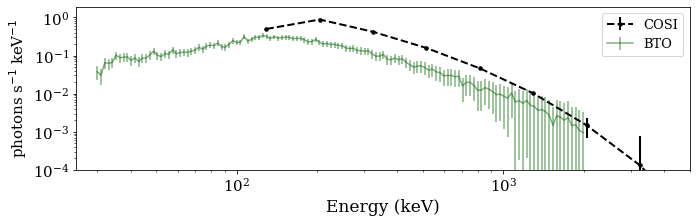}
\end{subfigure}
\caption{
\label{fig:lGRB_LC} [top] The lightcurve for a simulated long GRB with parameters listed in Table \ref{tab:GRB_params}. [bottom] The spectral response for the simulated GRB in units of ph/s/keV for COSI (dashed black line) and BTO (green line). The COSI spectrum covers 100 keV to 10 MeV and includes 10 energy bins. The COSI binning is set by computational limitations associated with calculating the current detector response matrix. In the future, the number of bins could increase. The BTO data covers 30 keV to 2 MeV with 200 energy bins, thus creating a finer sampling of the lGRB's spectral structure. The count rates for COSI and BTO become very low past 5 MeV and 1 MeV, respectively. However, spectral information up to 10 MeV and 2 MeV for COSI and BTO, respectively, will be recorded and used for fitting. The errors are given by Poisson statistics. }
\end{figure}

Figure \ref{fig:lGRB_LC} shows the lightcurve (top) and the corresponding BTO and COSI spectra (bottom) for a simulated  lGRB event directly above the COSI/BTO instruments (i.e. on-axis). The parameters for the GRB are included in Table \ref{tab:GRB_params}. The COSI spectrum is converted from units of photons/cm$^{-2}$/s/keV to units of photons/s/keV using the instrument's effective area over the 100 keV to 5 MeV bandpass (Neights et al. in prep). With an effective area peaking at $\sim$30 cm$^2$ from $\sim$200--500 keV \cite{tomsick2023}, COSI has a 
better response than the BTO detectors and achieves good statistics to higher energies. In the case of a typical lGRB, COSI provides well constrained data out to $\sim$5 MeV. BTO, on the other hand, has a sufficient response down to 30 keV and extends the observed bandpass to lower energies to capture the energy peak turnover at $\sim$100 keV. BTO is limited at higher energies, providing good statistics only up to $\sim$1 MeV. Unlike COSI, BTO will return event-by-event data for triggered events in the 30 keV to 10 MeV range. Therefore, the BTO spectrum can utilize finer bins, while event count rates allow.

\begin{figure}
\centering
\begin{subfigure}[b]{0.9\textwidth}
   \includegraphics[width=1\linewidth]{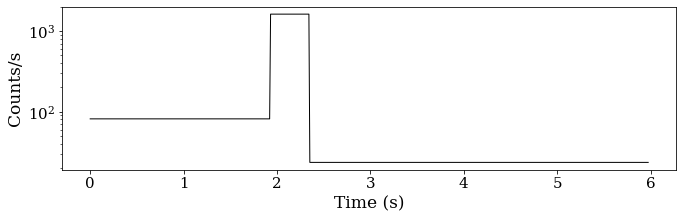} 
\end{subfigure}

\begin{subfigure}[b]{0.9\textwidth}
   \includegraphics[width=1\linewidth]{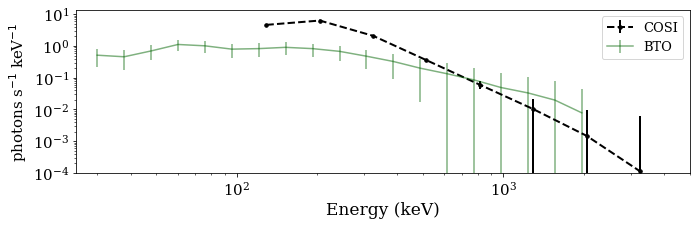}
\end{subfigure}
\caption{
\label{fig:sGRB_LC} [top] The lightcurve for a simulated short GRB with parameters listed in Table \ref{tab:GRB_params}. [bottom] The spectral response for the simulated GRB in units of ph/s/keV for COSI (dotted black line) and BTO (green line). The binning for the COSI data is the same as in Figure \ref{fig:lGRB_LC}. The BTO data uses 25 energy bins. The errors assume Poisson statistics.}
\end{figure}

Figure \ref{fig:sGRB_LC} shows the lightcurve (top) and the corresponding BTO and COSI spectra (bottom) for a simulated sGRB at 70$\degree$ off-axis from the COSI/BTO instrument zenith. The parameters for the GRB are included in Table \ref{tab:GRB_params}. The COSI and BTO spectra includes 10 and 25 energy bins, respectively. Similar to the lGRB, BTO is capable of providing well constrained data at energies lower than 100 keV but becomes dominated by errors closer to $\sim$600--700 keV. This is due to the sGRB having less total counts than the lGRB. COSI provides good statistics past 1 MeV.

\begin{table}[t!]
\caption{Simulated GRB Parameters} 
\label{tab:GRB_params}
\begin{center}       
\begin{tabular}{l|l|l}
\hline
\rule[-1ex]{0pt}{3.5ex}  Property & Long GRB & Short GRB  \\
\hline\hline
Spectrum Type & Band & Comptonized \\
\hspace{3mm}\textit{Parameters} & \hspace{3mm} E$_{break}$ = 841.8 & \hspace{3mm} E$_{peak}$ = 1482.6 \\
& \hspace{3mm} $\alpha$ = -1.02 & \hspace{3mm} $\alpha$ = -0.71 \\
& \hspace{3mm} $\beta$ = -2.83 & \hspace{3mm} --- \\
Duration (s) & 40.6 & 0.41 \\
Average Flux [ph/cm$^2$/s] & 10.72 & 35.38 \\
\hline 
\end{tabular}
\end{center} 
\end{table}

\subsection{Magnetar Flares}\label{subsec:magnetar}

There are two types of gamma-ray signals originating from magnetars: soft gamma-ray repeaters (SGRs) and giant flares (GFs). SGRs occur in young, slowly rotating magnetars and appear as dozens of rapid, repeating soft gamma-ray signals \cite{Thompson1996}. SGRs have a known initial hard transient phase of 0.2s \cite{Duncan1992, Thompson1995}. About once every 50 years \cite{Palmer2005}, an SGR's host magnetar will emit a GF with an energy $>$10$^3$ times the energy of the typical SGR burst in less than one second \cite{Mazets1979, Fenimore1996, Hurley1999, Mereghetti2024}. This prompt emission often peaks at $>$10--100s of keV \cite{Mereghetti2024, Kaspi2017, Boggs2007} and is well constrained in the 20 keV to 1 MeV band. The prompt GF emission is followed by a softer emission tail spanning several minutes and characterized by a periodic modulation intrinsic to the neutron star's rotation \cite{Mereghetti2024}. There are nearly 30 SGR catalogued in the Milky Way Galaxy, however very few have exhibited GFs. The few confirmed GFs include SGR 0526-66 \cite{Mazets1979, Cline1980, Mazets1982}, SGR 1900+14 \cite{Hurley1999, Ferroci1999, Mazets1999}, and SGR 1806-20 \cite{Hurley2005, Boggs2007, Palmer2005,Frederiks2007}. GFs outside of the galaxy are difficult to distinguish from sGRB, though it is possible to identify extragalactic GF through the event's spectral and timing properties as well as a lack of gravitational waves \cite{Mereghetti2024, Minaev2024}.

BTO's bandpass and large FOV  make it a great instrument for detecting rare, extragalactic events which peak in the soft gamma-ray regime---such as magnetar GFs. Additionally BTO's smaller effective area (when compared to COSI or GBM) is advantageous when it comes to magnetar GFs within the Galaxy. This is due to a GF's brief but incredibly energetic prompt emission being strong enough to overwhelm gamma-ray instruments, thus temporarily blinding them. With a smaller effective area, BTO will be blinded for less time and therefore capable of monitoring the prompt emission decay and subsequent modulating emission tail. Simulations of Galactic GF with BTO and COSI are ongoing.

%The prompt emission from a GF within the Milky Way Galaxy would over saturate both the COSI and BTO instruments, blinding them to the first second of the burst. However, the longer-duration tail would be detected by both instruments. For BTO, .... The simulations for COSI magnetar detections are ongoing.

\begin{figure}
\begin{center}
\begin{tabular}{c}
\includegraphics[height=5cm]{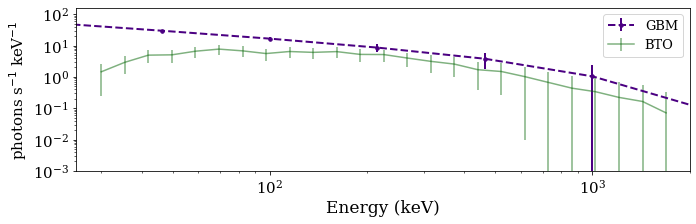}
\end{tabular}
\end{center}
\caption 
{ \label{fig:mag_GF}
The spectra for GRB231115A---a magnetar giant flare event in M82. The GBM spectrum (purple) is calculated from the fit to the observed GBM spectrum performed by \cite{Minaev2024}. The BTO spectrum (green) is simulated based off the GBM fit parameters. Poisson statistics are used to calculate the error bars on each energy bin. } 
\end{figure}

While unable to observe the prompt emission from Galactic magnetar GFs, both COSI and BTO will be capable of observing prompt emission from extragalactic giant flares. Using the spectral information from \cite{Mereghetti2024} and \cite{Minaev2024} for GRB231115A---a high probability extragalactic magnetar GF---we simulate BTO's spectral response for a similar GF at the instrument's zenith. Figure \ref{fig:mag_GF} shows the event's observed spectrum from GBM \cite{Minaev2024} and the simulated BTO spectrum. The observed GBM spectrum was fit over the time interval -0.02 s to 0.05 s with a comptonized model. The best fit parameters for this model are: normalization coefficient $A$ = 6.27$\pm0.7$ photons/cm$^2$/s/keV, a photon index of $\alpha$ = -0.34$\pm$0.2, and a peak energy $E_p$ = 637$\pm \sim70$ keV. The GBM spectrum was converted from photons/cm$^2$/s/keV to photons/s/keV using the GBM effective area curve fit in \cite{Smith2019}. The BTO spectrum was simulated with \textbf{MEGAlib} using the best fit parameters from GBM in the comptonized model and a fluence of 7.25e-7 erg/cm$^2$ in the 10--1000 keV range \cite{Minaev2024}. The simulations for COSI magnetar detections are ongoing.

%Extragalactic GF are visible to existing instruments to $\sim$10s of Mpc due to their luminosities exceeding $>$10$^{46}$ erg/s. However, the GF luminosity peak lasts fractions of a second before decreasing to 10$^{42}$ erg/s in the tail emission.

%The large FOV is good for this and energy range

\subsection{Terrestrial Gamma-ray Flashes}\label{subsec:TGF}

TGFs are rapid and intense bursts of gamma-rays originating in the Earth's atmosphere. Thunderstorms---and specifically lightning activity---are thought to be the progenitors of these events. However, the acceleration mechanism behind TGFs is not well understood \cite{Briggs2010, Dwyer2012}. While largely studied from the ground \cite{Belz2017}, TGFs are also detected by space-based gamma-ray telescopes such as the Reuven Ramaty High Energy Solar Spectroscopic Imager \cite[RHESSI]{Smith2005, Grefenstette2009}, GBM \cite{Briggs2010}, AGILE \cite{Marisaldi2010, Marisaldi2015}, and were even discovered by BATSE in 1994 \cite{Fishman1994}. 

TGF signals are very brief, with a single TGF pulse lasting less than a millisecond. A TGF event typically consists of a single pulse with energies extending up to 25 MeV, though events have been reported up to 38 MeV \cite{Briggs2010}. TGF emission is well modeled by bremsstrahlung radiation emitted by electrons accelerated in the relativistic runaway electron avalanche (RREA) process \cite{Gurevich1992}. Due to COSI/BTO's $\sim$0$\degree $\footnote{The current requirement for the COSI spacecraft is an orbit with $<$2$\degree$ inclination} inclination orbit, the instruments will follow the Earth's equator and spend periods of time over tropical regions where storms frequently occur thus increasing detection probabilities. However, great timing resolution ($<$0.5 ms) is needed to detect these events. BTO's event-by-event mode satisfies this requirement. 

%Thus increasing the chances of detecting a TGF. At an altitude of QQQ.

We simulate the BTO instrument response to a TGF event using \textbf{MEGAlib}. We build a TGF spectrum using the model from \cite{Lindanger2021}:

\begin{equation*}
    f(E) = \frac{1}{E} \exp{\frac{-E}{7.3 \mathrm{MeV}}}
\end{equation*}

where $E$ is the incoming photon energy in MeV. The spectrum is calculated from 30 keV to 40 MeV. The TGF photons are then beamed upward with a Gaussian beaming profile with a half angle of 30$\degree$ and Gaussian 1$\sigma$ value of 18$\degree$ as determined from Appendix B of \cite{Lindanger2021}. To simulate TGF events occurring on Earth, the TGF spectrum is passed through an atmospheric model created with the Python package \textbf{pyatmo}\footnote{Website: \url{https://github.com/jabesq-org/pyatmo}}. The resulting spectrum contains information on atmospheric interactions and is saved as the ``post-atmosphere" TGF spectrum. The BTO detector response is then simulated by ``observing" the post-atmosphere TGF spectrum with the BTO detector geometry. 
%The post-atmosphere spectrum is then ``observed" with the BTO detector geometry from 30 keV to 2 MeV to simulate the BTO detector response.
%An event duration of 0.001 seconds was used for the TGF \cite{Lindanger2021}. The TGF source flux was normalized by the number of injected photons.
%to have $\sim$250 counts per millisecond using the photons/s reported for different TGF events in \cite{Lindanger2021}.

\begin{figure}
\begin{center}
\begin{tabular}{c}
\includegraphics[height=7cm]{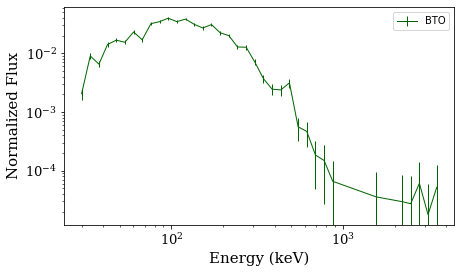}
\end{tabular}
\end{center}
\caption 
{ \label{fig:TGF_spec}
A TGF spectrum as observed by BTO. The spectrum is calculated for a 1 ms TGF at an initial altitude of 10 km. The flux is normalized by the number of photons injected into the simulation (1e7 photons).} 
\end{figure}

Figure \ref{fig:TGF_spec} shows the normalized TGF spectrum as observed by a single BTO detector for a TGF event with a starting altitude of 10 km and a 0.001 s duration \cite{Lindanger2021}. The spectrum is calculated from 30 keV to 2 MeV range and separated into 25 energy bins. The TGF source flux is normalized by the number of injected photons. With $>$20 energy bins, BTO is able to constrain the 511 keV peak produced by the TGF's gamma-rays inducing electron--positron annihilation in the Earth's atmosphere \cite{Kochkin2018, Wada2019}. The BTO spectrum provides good statistics up to $\sim$1 MeV for this TGF event.

Nominally, the main COSI detectors will not observe TGF events due to the anti-coincidence requirement between the COSI detectors and active shields. Since TGFs come from Earth (i.e. below the instrument), photons from these events will always pass through the active BGO shields before reaching the COSI detectors, and will therefore, be removed from the final, downlinked data products. An exception may occur when COSI points toward the Crab Nebula for calibrations, during which the telescope will be angled in such a way that a TGF event could enter the main COSI FOV without passing through a shield first. While the shields will save lightcurve information, the timing resolution in the shields (50 ms) is not suitable for detecting TGF events.

%However, the BGO shields will produce lightcurves in two energy regimes that may be used for rough TGF analysis. The lightcurves will cover a high-gain and low-gain regime with bandpasses of 80 keV to 2 MeV and 2 MeV to 10 MeV, respectively. Both lightcurve datasets will have a time resolution of 50 ms

\section{Conclusions}\label{sec:conclusions}

In this paper, we discuss initial calculations for the expected science returnables from the Background and Transient Observer (BTO) and the compatibility of the BTO and Compton Spectrometer and Imager (COSI) instruments. COSI is a wide-field gamma-ray telescope that will launch as a SMEX in 2027 and provide imaging, spectroscopy, and polarization measurements in the 0.2 to 5 MeV range. In addition to the main instrument, COSI will house the BTO detector system which will provide spectral information in the 0.03 MeV to 2 MeV bandpass. With overlapping bandpasses and FOVs, BTO and COSI enable spectral analysis across a broad gamma-ray bandpass. The combination of spectral information from these two instruments will enable:

\begin{itemize}
  \item Localization of transient gamma-ray sources within the shared COSI and BTO FOVs to $\sim$4$\degree$ (via COSI)
  \item Measurements of the energy peak turnover and photospheric or non-thermal emission components in GRB
  \item Cataloging of magnetar events including their characteristic timescales, temperatures, and brightness profiles
  \item Identification of thunderstorms which produce TGF events
\end{itemize}

Simulations of lGRB, sGRB, magnetar GFs, and TGFs are performed to showcase BTO's response to these transient events. In the case of GRBs, the expected detection rates are also calculated using the BATSE GRB catalogue. Using the BTO geometry over a 2 year nominal mission, we expect to detect $\sim$295 GRB events with $>$5$\sigma$ confidence, though this number might be slightly overestimated (by $\sim$50 GRB/year) when comparing with the GBM detection rate. Approximately 10\% of these events will be short GRB, resulting in $\sim$20 sGRB detections over the 2 year nominal mission duration. This will be a sufficient sample to measure time lags between sGRB events and their corresponding gravitational wave signals.

For magnetar GFs and TGFs, we simulate spectra from BTO using models from the literature. In both cases, BTO has good statistics up to $\sim$1 MeV where the data becomes error dominated. The magnetar data are compared to an observed GBM spectrum for an extragalactic GF in M82. Further analysis on the event rates of magnetar GFs as well as magnetar SGRs and TGFs is in progress.

%Outline

%1. Introduction

% 2. Mission Overview

% 2.1 The COSI Instrument

% 2.2 The BTO Instrument
%OVERLAPPING FOV
%OVERLAPPING ENERGIES

% 3. Gamma-Ray Bursts
%    Figure: GRB spectrum from COSI and BTO in same plot

% 4. Magnetar Flares
%    Figure: magnetar spectrum from COSI and BTO in same plot

% 5. TGFs
%    Figure: just BTO

% 6. Conclusions

\acknowledgments % equivalent to \section*{ACKNOWLEDGMENTS}       
 
The authors would like to acknowledge Alex Lowell (SSL), Brent Mochizuki (SSL), Josh Forgione (SSL), and Greg Dalton (SSL) for their support and guidance in developing the BTO instrument.

COSI and BTO are a NASA Small Explorer mission and Student collaboration project supported under NASA contract 80GSFC21C0059. This material is based upon work supported by the National Science Foundation Graduate Research Fellowship under Grant No. DGE 2146752 and support from the H2H8 organization. 

% References

%FOR SPIE
\printbibliography

%\bibliography{report} % bibliography data in report.bib
%\bibliographystyle{jwapjbib} % makes bibtex use spiebib.bst
%\bibliographystyle{spiebib}

%\input{main.bbl}

\end{document}